\documentclass[conference]{IEEEtran}
\usepackage{graphicx, array, amssymb, psfrag, graphics, setspace, stfloats}
\usepackage{amsfonts}
\usepackage[caption=false,font=normalsize,labelfont=sf,textfont=sf]{subfig}
\usepackage[toc,abbreviations,nonumberlist,nogroupskip,nopostdot]{glossaries-extra}
\glssetcategoryattribute{general}{glossdesc}{firstuc}
\glssetcategoryattribute{abbreviation}{glossdesc}{title}
\usepackage{mathtools}
\usepackage{multirow}
\usepackage{amsmath}
\usepackage{bibentry}

\DeclareMathOperator*{\argmin}{arg\,min}

\newcommand{\m}[1]{\mathbf{#1}}
\newcommand{\E}[1]{\mathbb{E} \left[{#1}\right]}
\newcommand{\Prob}[1]{\text{Pr} \left({#1}\right)}

\newcommand{\myabs}[1]{\left\lvert#1\right\rvert}

\def\scalingFig{0.3}

\newglossaryentry{computer}
{
name=computer,
description={A programmable machine that receives input data,
  stores and manipulates the data, and provides
  formatted output}
}

\newglossary*{nomenclature}{Nomenclature}

\newabbreviation{ISI}{ISI}					{inter-symbol interference}
\newabbreviation{DAC}{DAC}					{digital-to-analog converter}
\newabbreviation{ADC}{ADC}					{analog-to-digital converter}
\newabbreviation{DSP}{DSP}					{digital signal processing}
\newabbreviation{TX}{TX}					{transmitter}
\newabbreviation{RX}{RX}					{receiver}
\newabbreviation{PSK}{PSK}					{phase shift keying}
\newabbreviation{QAM}{QAM}					{quadrature-amplitude modulation}
\newabbreviation{FEC}{FEC}					{forward error correction}
\newabbreviation{SOP}{SOP}					{state-of-polarization}
\newabbreviation{FF}{FF}					{feed-forward}
\newabbreviation{FFEQ}{FFEQ}				{feed-forward equalizer}
\newabbreviation{BER}{BER}					{bit error rate}
\newabbreviation{SNR}{SNR}					{signal-to-noise ratio}
\newabbreviation{RSNR}{RSNR}				{required SNR}
\newabbreviation{SNDR}{SNDR}				{signal-to-noise-and-distortion ratio}
\newabbreviation{SFDR}{SFDR}				{spurious free dynamic range}
\newabbreviation{RC}{RC}					{raised cosine}
\newabbreviation{RRC}{RRC}					{root raised cosine}
\newabbreviation{ENOB}{ENOB}				{effective number of bits}
\newabbreviation{GD}{GD}					{group delay}
\newabbreviation{CMD}{CMD}					{chromatic dispersion}
\newabbreviation{PMD}{PMD}					{polarization mode dispersion}
\newabbreviation{PDL}{PDL}					{polarization dependent loss}
\newabbreviation{ASE}{ASE}					{amplified spontaneous emission}
\newabbreviation{LMS}{LMS}					{least mean squares}
\newabbreviation{APA}{APA}					{affine projection algorithm}
\newabbreviation{NLMS}{NLMS}				{normalized LMS}
\newabbreviation{MMSE}{MMSE}				{minimum mean square error}
\newabbreviation{CMA}{CMA}					{Constant modulus algorithm}
\newabbreviation{RLS}{RLS}					{recursive least squares}
\newabbreviation{LS}{LS}					{least squares}
\newabbreviation{LO}{LO}					{local-oscillator}
\newabbreviation{CR}{CR}					{carrier-recovery}
\newabbreviation{ASIC}{ASIC}				{application-specific integrated circuits}
\newabbreviation{FIR}{FIR}					{finite impulse response}
\newabbreviation{DD-LMS}{DD-LMS}			{decision-directed least mean squares}
\newabbreviation{DD}{DD}					{decision-directed}
\newabbreviation{CS-DAC}{CS-DAC}			{current-steering DAC}
\newabbreviation{LSB}{LSB}					{least-significant bit}
\newabbreviation{MSB}{MSB}					{most-significant bit}
\newabbreviation{DNL}{DNL}					{differential non-linearity}
\newabbreviation{INL}{INL}					{integral non-linearity}
\newabbreviation{DGD}{DGD}					{differential group delay}
\newabbreviation{FFT}{FFT}					{fast-Fourier transform}
\newabbreviation{IFFT}{IFFT}				{inverse fast-Fourier transform}
\newabbreviation{DFT}{DFT}					{discrete Fourier transform}
\newabbreviation{IDFT}{IDFT}				{inverse discrete Fourier transform}
\newabbreviation{FT}{FT}					{Fourier transform}
\newabbreviation{MSE}{MSE}					{mean square error}
\newabbreviation{HD}{HD}					{hard decision}
\newabbreviation{SD}{SD}					{soft decision}
\newabbreviation{LDPC}{LDPC}				{low density parity check}
\newabbreviation{CW}{CW}					{continuous wave}
\newabbreviation{PBC}{PBC}					{polarization beam combiner}
\newabbreviation{MIMO}{MIMO}				{multiple-input and multiple-output}
\newabbreviation{OPGW}{OPGW}				{optical ground wire}
\newabbreviation{ZF}{ZF}					{zero-forcing}
\newabbreviation{CAZAC}{CAZAC}				{Constant amplitude zero auto-correlation}
\newabbreviation{CFO}{CFO}					{carrier frequency offset}
\newabbreviation{MA}{MA}					{moving average}
\newabbreviation{DE}{DE}					{Differential evolution}
\newabbreviation{SA}{SA}					{simulated annealing}
\newabbreviation{DEM}{DEM}					{Dynamic element matching}
\newabbreviation{LUT}{LUT}					{lookup table}
\newabbreviation{DP}{DP}					{dynamic programming}
\newabbreviation{DPC}{DPC}					{digital pre-compensation}
\newabbreviation{NN}{NN}					{neural network}
\newabbreviation{MLSE}{MLSE}				{maximum–likelihood sequence estimation}
\newabbreviation{LE}{LE}					{linear equalizer}
\newabbreviation{DFE}{DFE}					{decision–feedback equalizer}
\newabbreviation{THP}{THP}					{Tomlinson-Harashima precoding}
\newabbreviation{HW}{HW}					{hardware}
\newabbreviation{PS}{PS}					{pilot sequence}
\newabbreviation{SW-LS}{SW-LS}				{sliding window least squares}
\newabbreviation{RD-Kalman}{RD-Kalman}      {radius-directed Kalman}	
\newabbreviation{RMS}{RMS}                  {root mean square}
\newabbreviation{SQNR}{SQNR}                {signal-to-quantization-noise ratio}
\newabbreviation{PDF}{PDF}                  {probability density function}
\newabbreviation{CDF}{CDF}                  {cumulative distribution function}
\newabbreviation{AWGN}{AWGN}			    {additive white Gaussian noise}

\newglossaryentry{dingledorf}
{
type=nomenclature,
name=dingledorf,
description={A person of supposed average intelligence who makes incredibly brainless misjudgments}
}

\newglossary*{symbols}{List of Symbols}
\newglossaryentry{rvec}
{
name={$\mathbf{v}$},
sort={label},
type=symbols,
description={Random vector: a location in n-dimensional Cartesian space, where each dimensional component is determined by a random process}
}
\newglossaryentry{C}
{
name={\ensuremath{\mathcal{C}}},
sort={label},
type=symbols,
description={Configuration space}
}
\newglossaryentry{C-space}
{
name={\ensuremath{\mathcal{C}\text{-space}}},
sort={label},
type=symbols,
description={Configuration space}
}
\newglossaryentry{C-free}
{
name={\ensuremath{\mathcal{C}^{\text{free}}}},
sort={label},
type=symbols,
description={Configuration space free of constraints/obstacles}
}
\newglossaryentry{P}
{
name={\ensuremath{\mathcal{P}}},
sort={label},
type=symbols,
description={Path planning space}
}
\newglossaryentry{P-constraint}
{
name={\ensuremath{\mathcal{P}^{\text{constraint}}}},
sort={label},
type=symbols,
description={Path planning space corresponding to constraints (including obstacles)}
}
\newglossaryentry{P-free}
{
name={\ensuremath{\mathcal{P}^{\text{free}}}},
sort={label},
type=symbols,
description={$\equiv \mathcal{P} \setminus \mathcal{P}^{\text{constraint}}$ Path planning space free of constraints/obstacles}
}
\newglossaryentry{p-goal}
{
name={\ensuremath{\textbf{p}^{\text{goal}}}},
sort={label},
type=symbols,
description={Destination point}
}
\newglossaryentry{p-start}
{
name={\ensuremath{\textbf{p}^{\text{start}}}},
sort={label},
type=symbols,
description={Starting point}
}
\newglossaryentry{W}
{
name={\ensuremath{\mathcal{W}}},
sort={label},
type=symbols,
description={Workspace}
}

\graphicspath{{Fig/}} 
\begin{document}

\title{Current-Steering DAC Architecture Design for Amplitude Mismatch Error Minimization}

\author{Ramin~Babaee, \textit{Student Member}, \textit{IEEE}, Shahab Oveis Gharan, and Martin Bouchard, \textit{Senior Member}, \textit{IEEE}\\
	\thanks{}}

 \maketitle

\begin{abstract}
We propose a novel \gls*{DAC} weighting architecture that statistically minimizes the distortion caused by random current mismatches. Unlike binary, thermometer-coded, and segmented DACs, the current weights of the proposed architecture are not an integer power of 2 or any other integer number. We present a heuristic algorithm for a static mapping of DAC input codewords into corresponding DAC switches. High-level Matlab simulations are performed to illustrate the static performance improvement over the segmented structure. 
\end{abstract}


\section{Introduction}
\IEEEPARstart{T}{he} rapid growth of high-speed communication systems requires the employment of faster analog components. Current steering DACs have been very successful in fulfilling the speed requirements due to their advantageous architecture designs \cite{9863991}. Amplitude and timing mismatch errors are the dominant factors in limiting the performance of these DACs. The static current mismatch degrades the signal at low frequencies and the timing errors distort the signal at higher frequencies. 

The most common examples of current-steering DACs are thermometer-coded, binary-weighted, and segmented \cite{5638288}. For an $N$-bit DAC, a thermometer-coded structure requires $2^N-1$ switches, each steering a current source of identical weight. Each \gls*{LSB} increment or decrement of digital input changes the state of only one switch. A decoding logic is needed to map the digital input words to thermometer codes. Although the circuit is complex to implement, it achieves guaranteed monotonicity. Contrary to thermometer-coded DACs, binary-weighted DACs have the smallest number of switches, i.e., $N$. The associated weights of current sources are in powers of $2$, i.e., $1$, $2$, ..., $2^{N-1}$. Binary DACs are not necessarily monotonic but are very efficient in area and power. Also, no decoding logic is needed as the input word bits directly control the state of switches. A segmented approach merges binary weighting for \gls*{LSB} bits and thermometer coding for \gls*{MSB} bits. This hybrid method provides flexibility for \gls*{DAC} design area and performance constraints \cite{5638288}. 
The redundancy offered by the unary cells enables the use of two popular distortion mitigation techniques, \gls*{DEM} and mapping techniques. \gls*{DEM} algorithms stochastically or deterministically randomize the selection of unary cells to improve the \gls*{SFDR} performance at the cost of incurring noise floor penalty \cite{RTSC,911479,476173,5420027,913021,9485113}. However, the smart selection solution offered by the mapping algorithms presents superior performance without affecting the noise floor. The mapping techniques largely use either amplitude errors or timing errors to find an optimized mapping of unary cells. Switching sequence optimization based on amplitude error minimization for systematic and graded errors is studied in \cite{75066,735536,808896,850417}. A sort-and-group method is usually utilized to find the most efficient selection.

In this study, we introduce a novel DAC architecture framework for improving the statistical DAC performance due to amplitude mismatches. We first develop a statistical metric that can be used for minimizing the random current mismatch errors. Utilizing the derived metric, a novel optimized architecture is obtained which can operate more efficiently compared to a commonly used segmented structure. We demonstrate though simulations that the optimized architecture has an improved statistical performance compared to segmentation and therefore improves the chip yield. 

The rest of the paper is organized as follows. In Section \ref{sec:be_analysis}, a detailed analysis of random amplitude errors is presented. Section \ref{sec:be_architecture} discusses the proposed DAC weighting scheme. Simulation results are provided in Section \ref{sec:simulaion_results} and followed by conclusions in Section \ref{sec:conclusion}.

\textbf{Notation:} Vectors and scalars are represented by bold letters and non-bold italic letters, respectively. $\m{B}_i$ represents the $i$-th element of vector $\m{B}$. $(\cdot)^\text{T}$ indicates transpose of a vector/matrix. Symbol $\E{\cdot}$ is used for statistical expectation operation. The cardinality of set $\mathcal{R}$ is also denoted by $|\mathcal{R}|$. 

\section{Amplitude Error Analysis}
\label{sec:be_analysis}

Let $N$ and $L$ denote the number of input bits and current switches of a DAC, respectively. The ideal weights of current sources are represented by an $L$-dimensional basis vector $\m{B}$. A current source of weight $\m{B}_i$ is usually implemented by $\m{B}_i$ parallel unit current sources. Note that $\m{B}_i = 2^i$ and $\m{B}_i =1$ for a binary DAC and a unary DAC, respectively. However, due to component mismatches, the ideal current weights are not achieved. 

Let the $L$-dimensional vector $\m{\Delta}$ denote the deviation from the ideal current values. The mismatch for current element $i$, i.e., $\m{\Delta}_i$, is statistically modeled as an uncorrelated Gaussian random variable with a standard deviation of $\sqrt{\m{B}}_i \sigma_\delta$, where $\sigma_\delta$ is the unit current source standard deviation. Note that even though the current mismatch is statistically modeled as a random variable, it has a static value for a given DAC sample. The input to the \gls*{DAC} is an integer number denoted by $x$ in the range $[ 0 \quad 2^N-1]$ and has a probability density function $\Prob{x}$. The binary representation of a given integer input $x$ is expressed by an $L$-dimensional binary vector $\m{W}(x)$ which satisfies $\m{W}^\text{T}(x)\m{B} = x$. The current output of the DAC for input word $x$ can be formulated as
\begin{equation}
y = \m{W}^\text{T}(x) (\m{B}+\m{\Delta}).
\end{equation}
Therefore, the signal at the receiver after the removal of the offset term can be expressed as
\begin{align}
z & = y - \mathbb{E}_y[y] \nonumber \\
& = \sum_{i=0}^{L-1} \m{W}_i(x)(\m{B}_i + \m{\Delta}_i) - \mathbb{E}_x \bigg[ \sum_{i=0}^{L-1}  \m{W}_i(x)(\m{B}_i + \m{\Delta}_i) \bigg] \nonumber  \\
& = \sum_{i=0}^{L-1} \m{W}_i(x)\m{B}_i +\sum_{i=0}^{L-1} \m{W}_i(x) \m{\Delta}_i \nonumber  \\
& \hspace{3ex} -  \mathbb{E}_x \bigg[ \sum_{i=0}^{L-1} \m{W}_i(x) \m{B}_i \bigg]  + \mathbb{E}_x \bigg[ \sum_{i=0}^{L-1} \m{W}_i(x) \m{\Delta}_i) \bigg]   \nonumber  \\
& = x - \mathbb{E}_x \big[x\big] + \sum_{i=0}^{L-1} \m{W}_i(x) \m{\Delta}_i - \sum_{i=0}^{L-1} \mathbb{E}_x \big[\m{W}_i(x) \m{\Delta}_i \big],
\end{align}
The error caused by the current mismatch at the receiver is then given by
\begin{align}
e_m(x) & = z-(x - \mathbb{E}_x) \nonumber \\ 
& = \sum_{i=0}^{L-1} \m{W}_i(x) \m{\Delta}_i -  \sum_{i=0}^{L-1} \mathbb{E}_x \big[\m{W}_i(x)\big] \m{\Delta}_i \nonumber \\
& = \sum_{i=0}^{L-1} \Big( \m{W}_i(x) - \mathbb{E}_x \big[\m{W}_i(x)\big] \Big) \m{\Delta}_i,
\label{eq:be_e}
\end{align}
where 
\begin{equation}
\mathbb{E}_x \big[\m{W}_i(x)\big] = \sum_{x=0}^{2^N-1} \Prob{x} \m{W}_i(x).
\end{equation}
The ensemble average of squared error, i.e., \gls*{MSE}, over DAC realizations and input $x$ can be computed as
\begin{align}
\mathbb{E}_{x,\m{\Delta}} \big[|e_m|^2\big] & = \mathbb{E}_{x,\m{\Delta}} \bigg[ \myabs{\sum_{i=0}^{L-1} \Big( \m{W}_i(x) - \mathbb{E}_x \big[\m{W}_i(x)\big] \Big)  \m{\Delta}_i}^2 \bigg] \nonumber \\ 
& = \sigma^2_\delta \; \mathbb{E}_x \bigg[ \sum_{i=0}^{L-1} \Big( \m{W}_i(x) - \mathbb{E}_x \big[\m{W}_i(x)\big] \Big)^2 \m{B}_i \bigg] \nonumber \\
& = \sigma^2_\delta \sum_{i=0}^{L-1} \Big( \mathbb{E}_x \big[ \m{W}^2_i(x) \big] - \mathbb{E}_x \big[ \m{W}_i(x)\big ]^2 \Big) \m{B}_i.
\end{align}
Since $\m{W}(x)$ is a binary vector, we have $\m{W}_i^2(x) = \m{W}_i(x)$. Therefore, 
\begin{equation}
\mathbb{E}_{x,\m{\Delta}} \big[|e_m|^2\big] = \sigma^2_\delta \sum_{i=0}^{L-1} \mathbb{E}_x \big[ \m{W}_i(x) \big] \big(1 - \mathbb{E}_x \big[ \m{W}_i(x) \big]\big) \m{B}_i.
\label{eq:be_stat_e2}
\end{equation}
If the \gls*{DAC} architecture offers redundancy, input $x$ might have multiple representation vectors. Let $\mathcal{R}(x)$ represent the set of all possible representations of $x$, i.e., 
\begin{equation}
\mathcal{R}(x) = \left\{\m{W} | \m{W}^\text{T} \m{B} = x\right\}.
\end{equation}
In this case, the representation used in Eq. (\ref{eq:be_stat_e2}) can influence the performance of the DAC. The optimum set of representations can be computed as the solution to the optimization problem
\begin{align}
\argmin_{\m{W}(0),...,\m{W}(2^N-1)} \mathbb{E}_{x,\m{\Delta}} \big[|e_m|^2\big]
\end{align}
which is expanded as
\begin{align}
\argmin_{\m{W}(0),...,\m{W}(2^N-1)} \sum_{i=0}^{L-1} \mathbb{E}_x \big[ \m{W}_i(x) \big] \big(1 - \mathbb{E}_x \big[ \m{W}_i(x) \big]\big) \m{B}_i.
\label{eq:mis_opt_rep}
\end{align}
This is a discrete non-convex optimization problem where one may try to find the optimal solution by exhausting all possible representations. Given that, on average there are $2^{L-N}$ representations for each value of $x$, computation complexity of the exhaustive search is $\mathcal{O}(2^{(L-N)2^{N}})$. Thus, this approach becomes computationally complex as $L$ grows. We use an iterative sub-optimal approach to solve the problem. 
A brief description of the iterative optimization method is outlined below.
\begin{itemize}
    \item For a given basis $\m{B}$, calculate all representations for each codeword.
    \item Initialize $\m{W}_{\text{opt}}(x)$ for  $x=0,...,2^N-1$ with a random representation.
    \item For each iteration of the algorithm, find the optimum representation for each codeword sequentially. This is achieved by fixing the representations of all values except $y$ and solving for
\begin{align}
\m{W}_\text{opt}(y) & = \argmin_{\m{W}(y) \in \mathcal{R}(y)} \sum_{i=0}^{L-1} \Big( \mathbb{E}_x  \big[ \m{W}_{\text{opt},i}(x) \big] \\ \nonumber
& \hspace{8ex} \big(1 - \mathbb{E}_x \big[ \m{W}_{\text{opt},i}(x) \big]\big) \m{B}_i \Big)
\end{align} using an exhaustive search.
    \item Repeat the search step for $T$ iterations or until $\m{W}_{\text{opt}}(y)$ remains the same for all values of $y$. 
\end{itemize}

\section{Weighting Scheme Design}
\label{sec:be_architecture}
One can use the derived metric in Eq. (\ref{eq:mis_opt_rep}) to find an over-complete basis that may attain improved performance compared to a segmented architecture. The optimization problem can be formulated as
\begin{align}
\m{B}_\text{opt} & = \argmin_{\m{B}} \Big( \argmin_{\substack{\m{W}(x,\m{B}) \\  0 \le x < 2^N}} \sum_{i=0}^{L-1} \big( \mathbb{E}_x \big[ \m{W}_i(x,\m{B}) \big] \nonumber \\
& \hspace{22ex} \big(1 - \mathbb{E}_x \big[ \m{W}_i(x,\m{B}) \big]\big) \m{B}_i \big) \Big).
\label{eq:be_opt_problem}
\end{align}
We have used the notation $\m{W}(x,\m{B})$ to highlight that the representation vectors are also a function of the basis. Eq. (\ref{eq:be_opt_problem}) is a non-convex integer non-linear programming. Therefore, a heuristic optimization algorithm must be used to solve it. 

From the many available techniques, we use \gls*{SA} \cite[Chapter 7]{optimization_book} for solving the optimization problem. A white Gaussian signal is assumed for the basis optimization where the \gls*{RMS} of the input signal is set to the optimum value that achieves maximum \gls*{SQNR} using simulation. This assumption determines the probability density function of input $x$, i.e., $\Prob{x}$. The reason for the Gaussian distribution assumption of the input is that a Gaussian-distributed signal is required at the channel input to achieve \gls*{AWGN} channel capacity in communications systems. Note that this optimization problem can be solved for any other distributions such as uniform input distribution. Alternatively, a sine wave signal can also be considered for this optimization. 

\begin{figure}[!tb]
\centering
\footnotesize
\includegraphics[scale=0.36]{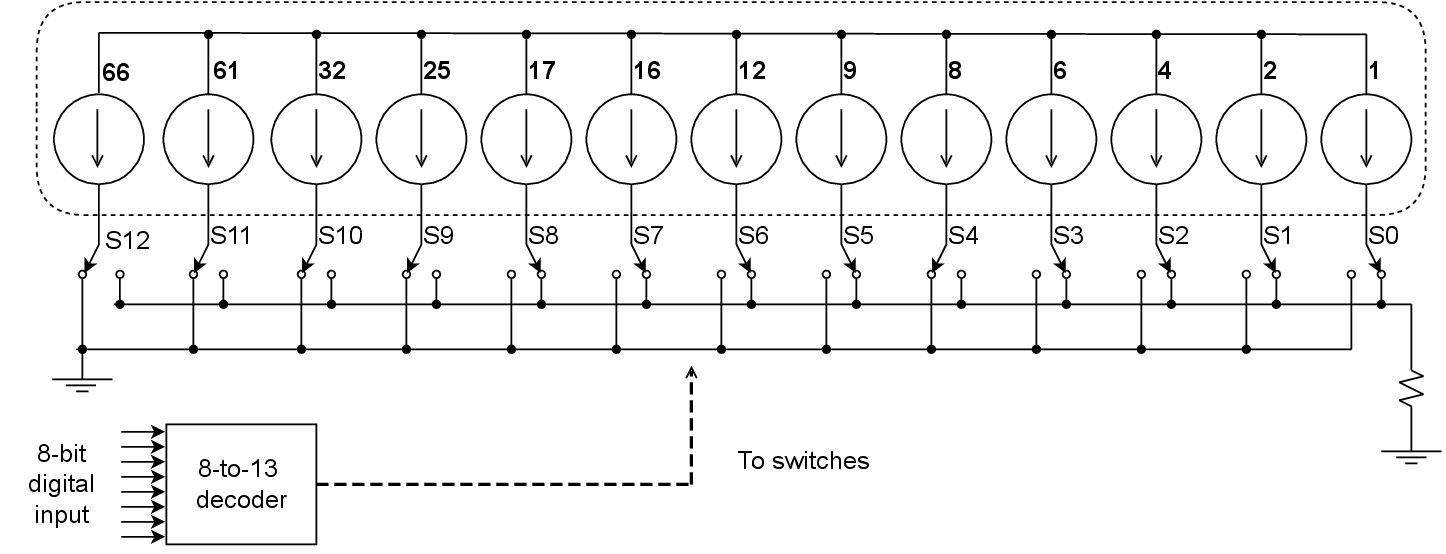}
\caption{Block diagram of the optimized 13 switches architecture for an 8-bit DAC.}
\label{fig:be_13_architecture}
\end{figure}

The \gls*{SA} optimization algorithm is run $100$ times and the best basis is selected from the results. It is worth mentioning that in each iteration of the \gls{SA} algorithm, the candidate basis must satisfy the completeness property, i.e., each integer number in the range $[0 \hspace{1ex} 2^N-1]$ can be formed as a binary combination of basis elements. Table \ref{tab:be_opt_basis} summarizes the optimized basis vectors for several values of $L$ for an 8-bit DAC. The optimal mappings of codewords 118 to 138 are presented in Table \ref{tab:be_single_rep} for the case of $L=13$. The block diagram of the optimized architecture for the case of 13 switches is also illustrated in Fig. \ref{fig:be_13_architecture}. Note that since the mapping of input codewords to DAC switches is static, it can be implemented using a $2^N$ \gls*{LUT} in the hardware.

\begin{table}[ht]
\footnotesize
\centering
\begin{tabular}{||l||*{13}{p{0.3em}}||}
\hline
\textbf{Basis Length} & \multicolumn{13}{l||}{\textbf{Optimized Basis}} \\ 
\hline
9  &  1&   2&   4&   8&  16&  32&  35&  77&  80&    &    &    &      \\
10 &  1&   2&   4&   8&  16&  17&  32&  33&  70&  72&    &    &      \\
11 &  1&   2&   4&   8&   8&  16&  17&  32&  33&  66&  70&    &      \\
12 &  1&   2&   4&   7&   8&  15&  15&  23&  25&  30&  61&  64&      \\
13 &  1&   2&   4&   6&   8&   9&  12&  16&  17&  25&  32&  61&  66  \\
\hline
\end{tabular}
\vspace{1ex}
\caption{Amplitude error optimized basis vectors for an 8-bit DAC.}
\label{tab:be_opt_basis}
\end{table}

\begin{table}[!tbh]
\footnotesize
\centering
\hspace{1ex}
\begin{tabular}{||l||*{13}{p{0.36em}}||}
\hline
$\m{B}$ 	  &1     &2     &4    &6     &8     &9    &12    &16    &17   &25    &32    &61    &66  \\
\hline\hline
   118     &0     &0     &0    &1     &0     &1    & 1    & 0    & 0   & 1    & 0    & 0    & 1  \\
   119     &1     &0     &0    &1     &0     &1    & 1    & 0    & 0   & 1    & 0    & 0    & 1  \\
   120     &0     &1     &0    &1     &0     &1    & 1    & 0    & 0   & 1    & 0    & 0    & 1  \\
   121     &1     &1     &0    &1     &0     &1    & 1    & 0    & 0   & 1    & 0    & 0    & 1  \\
   122     &0     &0     &1    &1     &0     &1    & 1    & 0    & 0   & 1    & 0    & 0    & 1  \\
   123     &1     &0     &1    &1     &0     &1    & 1    & 0    & 0   & 1    & 0    & 0    & 1  \\
   124     &0     &1     &1    &1     &0     &1    & 1    & 0    & 0   & 1    & 0    & 0    & 1  \\
   125     &1     &1     &1    &1     &0     &1    & 1    & 0    & 0   & 1    & 0    & 0    & 1  \\
   126     &0     &0     &0    &1     &0     &0    & 1    & 0    & 1   & 1    & 0    & 0    & 1  \\
   127     &1     &0     &0    &1     &0     &0    & 1    & 0    & 1   & 1    & 0    & 0    & 1  \\
   128     &0     &1     &0    &1     &0     &0    & 1    & 0    & 1   & 1    & 0    & 0    & 1  \\
   129     &1     &1     &0    &1     &0     &0    & 1    & 0    & 1   & 1    & 0    & 0    & 1  \\
   130     &0     &0     &1    &1     &0     &0    & 1    & 0    & 1   & 1    & 0    & 0    & 1  \\
   131     &1     &0     &1    &1     &0     &0    & 1    & 0    & 1   & 1    & 0    & 0    & 1  \\
   132     &0     &1     &1    &1     &0     &0    & 1    & 0    & 1   & 1    & 0    & 0    & 1  \\
   133     &1     &1     &1    &1     &0     &0    & 1    & 0    & 1   & 1    & 0    & 0    & 1  \\
   134     &1     &0     &1    &0     &0     &1    & 1    & 0    & 1   & 1    & 0    & 0    & 1  \\
   135     &0     &0     &0    &1     &0     &1    & 1    & 0    & 1   & 1    & 0    & 0    & 1  \\
   136     &1     &0     &0    &1     &0     &1    & 1    & 0    & 1   & 1    & 0    & 0    & 1  \\
   137     &0     &1     &0    &1     &0     &1    & 1    & 0    & 1   & 1    & 0    & 0    & 1  \\
   138     &1     &1     &0    &1     &0     &1    & 1    & 0    & 1   & 1    & 0    & 0    & 1  \\
\hline
\end{tabular}
\vspace{2ex}
\caption{Optimal representation of codewords $118$ to $138$ for the optimized basis of length $13$.}
\label{tab:be_single_rep}
\end{table}

\section{Simulation Results}
\label{sec:simulaion_results}

In this section, simulation results are presented to evaluate the performance of the proposed strategies and to validate our theoretical analysis. A high-level 8-bit \gls*{DAC} is simulated in MATLAB. The only impairment considered is current mismatches which are assumed to be independent and have a Gaussian distribution. The mean \gls*{SNDR} values are obtained by averaging over 100,000 DAC realizations.

\begin{figure}[!tb]
\centering
\tiny
\psfrag{Hybrid, best_next}{Segmented architecture}
\psfrag{Optimized, single_rep_opt}{Optimized architecture}
\psfrag{Basis length}{Basis length $L$}
\psfrag{Normalized cost (Analytical)}{\hspace{10ex} Normalized metric}
\includegraphics[scale=\scalingFig]{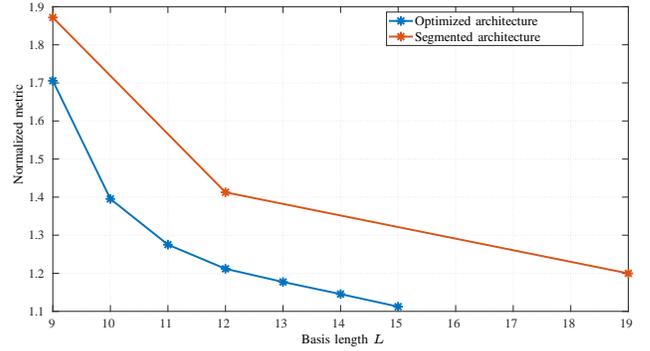}
\caption{Normalized statistical amplitude mismatch metric (Eq. (\ref{eq:be_stat_e2})) as a function of basis length $L$.}
\label{fig:be_metric}
\end{figure}

Fig. \ref{fig:be_metric} illustrates the derived statistical metric in Eq. (\ref{eq:be_stat_e2}) for a segmented \gls*{DAC} and the optimized bases. The $x$-axis is the number of \gls*{DAC} switches and the $y$-axis is the metric normalized by the metric of a fully thermometer-coded design. The three segmented DACs on the red curve are 2T+6B, 3T+5B, and 4T+4B architectures, where the first number represents the number of \gls*{MSB}s used for thermometer weighting and the second number denotes the number of binary bits Expectedly, as the number of switch elements increases, the performance improves. As evident in the figure, the optimized architecture outperforms the 4+4 segmentation, which requires 19 switches, with only 13 switches, corresponding to more than $30\%$ area saving. 

\begin{figure}[!tb]
\centering
\tiny
\psfrag{Optimized20, single_rep}{Optimized architecture}
\psfrag{Hybrid20, single_rep}{Segmented architecture}
\psfrag{SDR [dB]}{SNDR [dB]}
\psfrag{basis_len}{Basis length $L$}
\includegraphics[scale=\scalingFig]{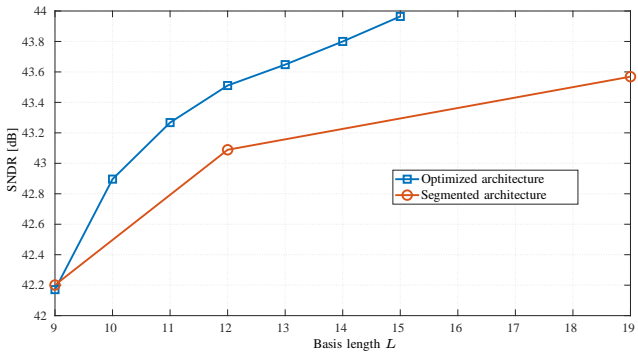}
\caption{Mean SNDR comparison of the proposed architecture and segmented DAC. The results are obtained by averaging over 100,000 realizations. }
\label{fig:be_mean_sndr}
\end{figure}

\begin{figure}[!tb]
\centering
\tiny
\psfrag{Optimized20, single_rep}{Optimized architecture}
\psfrag{Hybrid20, single_rep}{Segmented architecture}
\psfrag{SDR [dB], Yield = 95}{\hspace{5ex} SNDR [dB]}
\psfrag{basis_len}{Basis length $L$}
\includegraphics[scale=\scalingFig]{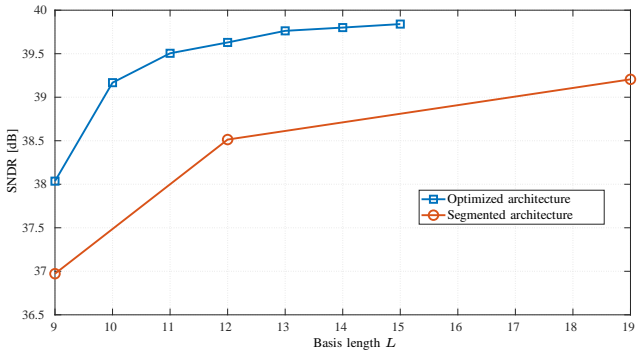}
\caption{SNDR 95th-percentile comparison of the proposed architecture and segmented DAC. The results are obtained from over 100,000 realizations.}
\label{fig:be_95_sndr}
\end{figure}

The \gls*{SNDR} due to component mismatches is plotted in Fig. \ref{fig:be_mean_sndr} for $\sigma_\delta=0.05$. The results confirm the superior performance of the proposed structure over the segmented architecture and are consistent with Fig. \ref{fig:be_metric}. The 95th-percentile \gls*{SNDR} is also depicted in Fig. \ref{fig:be_95_sndr} for comparison. It shows a significant benefit of the proposed weighting scheme over segmented DAC. The proposed optimized architecture not only improves the mean SNDR, but also significantly improves the 95th-percentile, which ultimately improves the chip yield. The superior performance is attained by using only 13 switches in the DAC, in contrast to the 19 switches required for the 4+4 segmentation.

\section{CONCLUSION}
\label{sec:conclusion}
We explored current mismatch errors caused by component mismatches in current-steering DACs. We derived the MSE of the amplitude mismatches and formulated an optimization framework to obtain a novel architecture that improves the \gls*{DAC} static performance compared to the traditional segmented structure. Simulation results demonstrated that the proposed architecture with only 13 switches outperforms segmented DAC with 19 switches, resulting in a significant area saving in addition to the improved chip yield. Although the paper focused on SNDR improvement, one can use the same framework for deriving superior architectures using other metrics such as \gls*{DNL}. In the next step of this research, we will conduct transistor-level simulations to verify the potential benefit of the proposed DAC architecture.

\bibliography{IEEEabrv,main}
\bibliographystyle{IEEEtran}

\end{document}